\documentclass[english,runningheads]{llncs}

\newif\ifwop
\wopfalse
\ifwop \usepackage[a4paper, showframe]{geometry} \usepackage{fullpage} \usepackage{lineno} \linenumbers \fi 
\usepackage[utf8]{inputenc}
\usepackage[T1]{fontenc}
\usepackage{babel}
\usepackage{hyperref}
\usepackage{csquotes,hyphenat,comment}
\usepackage{algorithm,algorithmicx,algpseudocode}
\usepackage{lineno}
\usepackage{thmtools}
\makeatletter
\let\c@author\relax
\makeatother
\usepackage[backend=biber]{biblatex}

\newcommand\digitalonly[1]{}
\usepackage{hyperref,xcolor}

\providecolor{Green}{HTML}{00a650}

\hypersetup{
	colorlinks   = true, %
	urlcolor     = blue, %
	linkcolor    = blue!50!red!78, %
	citecolor   = Green %
}

\AtEveryBibitem{ \clearlist {publisher} \clearfield {isbn} \clearfield{location} \clearfield{booktitle}
	\clearname {editor} \clearfield {series} \clearfield {url} \clearfield {issn} }

\usepackage{amssymb,mathtools} %
\usepackage{dsfont,interval,braket}
\usepackage[alignedforall]{lpform}
\usepackage{xparse,enumerate,needspace}
\usepackage{todonotes} 
\usepackage{tikz,subcaption,placeins} \usetikzlibrary{arrows,decorations.pathreplacing,patterns}
\usepackage{wrapfig}

\providecommand\digitalonly[1]{#1}
\RequirePackage{etoolbox,xspace}
\newcommand\threefield[3]{\ifmmode#1|#2|#3\else $#1|\allowbreak#2|\allowbreak#3$\xspace\fi}

\newcommand\hsym[1]{\ifmmode{#1}\else\ensuremath{#1}\xspace\fi}
\renewcommand*\do[1]{%
	\expandafter\newcommand\csname C#1\endcsname{\hsym{C_{\mathrm{#1}}}}%
}
\docsvlist{max,min,envy,idle}

\newcommand\QCmax[1][]{\threefield Q {#1} {\Cmax}}
\newcommand\QCmin[1][]{\threefield Q {#1} {\Cmin}}
\newcommand\QCenvy[1][]{\threefield Q {#1} {\Cenvy}}
\newcommand\QCidle[1][]{\threefield Q {#1} {\Cidle}}

\newcommand\pmax{\hsym{p_{\mathrm{max}}}}
\newcommand\smax{\hsym{s_{\mathrm{max}}}}
\newcommand\pmin{\hsym{p_{\mathrm{min}}}}

\newcommand{\inv}{{-1}}

\newcommand\HM[1][m,n]{\ifmmode\mathit{HM}(#1)\else\ensuremath{\HM[#1]}\xspace\fi}
\newcommand\imax{i_{\mathrm{max}}}
\newcommand\RAP{\ifmmode p_{ij} {\in} \sing{p_j, \infty}\else\ensuremath{\RAP}\xspace\fi}
\RequirePackage{interval}
\intervalconfig{separator symbol = {;}}
\providecommand\linterval[2][]{\interval[open left , #1]{#2}}

\providecommand\lInterval[2][]{\Interval[open left , #1]{#2}}

\newcommand\Interval[3][]{\interval[#1, separator symbol={\,..\,}]{#2}{#3}}
\newcommand\iv[1]{[{#1}]}
\usepackage{xparse,etoolbox,pgfgantt}
	\newcounter{currenttime}
	\newcounter{tmpcounter}
	\newcommand\addtask[1]{
		\setcounter{tmpcounter}{\value{currenttime}}
		\addtocounter{currenttime}{#1}
		\addtocounter{currenttime}{-1}
		\ganttbar[inline]
			{\ensuremath{#1}}
			{\value{tmpcounter}}
			{\value{currenttime}}
		\stepcounter{currenttime}
	}
	\DeclareListParser{\dolist}{ }
	\newcommand\addtasks[1]{
		\renewcommand\do[1]{\addtask {##1}}
		\dolist{#1}
	}
	\NewDocumentCommand{\newmachine}{o}{
		\setcounter{currenttime}{0}
		\IfNoValueF{#1}{\ganttbar[.style = {draw=none}]{#1}{0}{0}}
	}
\RequirePackage{dsfont}
\NewDocumentCommand\onevec{o}{%
	\IfValueTF{#1}{%
		\mathds 1_{#1}%
	}{%
		\mathds 1%
	}%
}
\NewDocumentCommand\zerovec{o}{%
	\IfValueTF{#1}{%
		\mathds O_{#1}%
	}{%
		\mathds O_{\N \times 1}%
	}%
}
\NewDocumentCommand\zero{o}{%
	\IfValueTF{#1}{\mathds O_{#1}}{\mathds O}%
}
\NewDocumentCommand\one{o}{%
	\IfValueTF{#1}{\mathds E_{#1}}{\mathds E}%
}

\usepackage{bbm}
\NewDocumentCommand\uvec{om}{%
	\IfValueTF{#1}{%
		\mathbbm e^{(#2)}_#1
	}{%
		\mathbbm e^{(#2)}
	}%
}
\RequirePackage{mathtools}
\DeclareMathOperator{\supp}{supp}
\DeclareMathOperator{\lcm}{lcm}

\DeclareMathOperator{\poly}{poly}

\DeclarePairedDelimiter\abs{\lvert}{\rvert}%
\DeclarePairedDelimiter\norm{\lVert}{\rVert}%
\DeclarePairedDelimiter\ceil{\lceil}{\rceil}
\DeclarePairedDelimiter\floor{\lfloor}{\rfloor}

\let\O=\relax
\DeclareMathOperator{\O}{\mathcal O}

\newcommand\OPT{\mathrm{OPT}}

\DeclarePairedDelimiter{\internalXenpar}{\lparen}{\rparen}
\NewDocumentCommand{\enpar}{s t~ m}{
	\IfBooleanTF{#2}{%
		{%
		\IfBooleanTF{#1}{\internalXenpar*{#3}}{\internalXenpar{#3}}%
		}%
	}{%
		\IfBooleanTF{#1}{\internalXenpar*{#3}}{\internalXenpar{#3}}%
	}%
}

\RequirePackage{dsfont}
\newcommand\N{\ensuremath{\mathds N}}
\newcommand\R{\ensuremath{\mathds R}}
\newcommand\Z{\ensuremath{\mathds Z}}

\newcommand\defeq\coloneqq
\newcommand\eqdef\eqqcolon
\newcommand\eps\varepsilon
\newcommand\super[2]{#2^{(#1)}}

\newcommand\sing[1]{\{#1\}}

\NewDocumentCommand{\semi}{t& t| t. O{;}}{%
	\IfBooleanTF#3{%
		.%
	}{%
		#4%
		\IfBooleanT{#1}{\text{\ and\ }}%
		\IfBooleanT{#2}{\text{\ or\ }}%
	}%
}%
	\makeatletter
	\newcommand{\dotminus}{\mathbin{\text{\@dotminus}}}
	\newcommand{\@dotminus}{%
		\ooalign{\hidewidth\raise1ex\hbox{.}\hidewidth\cr$\m@th-$\cr}%
	}
	\makeatother
{
\let\olddotminus=\dotminus
\renewcommand\dotminus{\mathrel{\olddotminus}}
}
\DeclareFontFamily{U}{mathx}{\hyphenchar\font45}
\DeclareFontShape{U}{mathx}{m}{n}{
	  <5> <6> <7> <8> <9> <10>
	  <10.95> <12> <14.4> <17.28> <20.74> <24.88>
	  mathx10
	  }{}
\DeclareSymbolFont{mathx}{U}{mathx}{m}{n}
\DeclareFontSubstitution{U}{mathx}{m}{n}
\DeclareMathAccent{\widecheck}{0}{mathx}{"71}
\DeclareMathAccent{\wideparen}{0}{mathx}{"75}
\newcommand\vecn\nu
\newcommand\vecm\mu
\newcommand\scalm{m}

\newcommand\mydelta\delta
\newcommand\mygamma\gamma
\newcommand\resch\rho
\newcommand\stopp{\,\hfill\ensuremath{\lrcorner}}
\providecommand\textcite[1]{\citet{#1}}
\newcommand\lipicsitem[1]{\textit{#1}}
\providecommand\appproof[1]{Proof}

\let\check=\widecheck
\let\hat=\widehat

\let\bullet=i

\NewDocumentCommand{\itemref}{s m}{%
	\IfBooleanTF#1{%
		(\ref*{#2})%
	}{%
		\hyperref[#2]{(\ref*{#2})}%
	}%
}

\providecommand\subref[2]{} \renewcommand\subref[2]{\hyperref[#1:#2]{\autoref*{#1}~\itemref*{#1:#2}}}
\newcommand\stmtref[1]{\hyperref[#1]{Statement \itemref*{#1}}}
\newcommand\dref[1]{\text{\hyperref[#1]{(\ref*{#1})}}}
\newcommand\dsubref[2]{\text{\hyperref[#1:#2]{(\ref*{#1}\ref*{#1:#2})}}}

\newcommand\subpoint[2][parfill]{%
	\csname #1\endcsname%
	\noindent%
	\textit{Proof of {#2}.}%
}

\NewDocumentCommand\overann{s O{\big} m m}{
	\IfBooleanTF{#1}{%
		\overset{\mathclap{#4}}{#3}%
	}{
		\overset{\mathclap{\substack{{#4} \\ #2\downarrow}}}{#3}%
	}
}
\NewDocumentCommand\underann{s O{\big} m m}{
	\IfBooleanTF{#1}{%
		\overset{\mathclap{#4}}{#3}%
	}{
		\underset{\mathclap{\substack{ #2\uparrow \\ {#4}}}}{#3}%
	}
}
\makeatletter
\xspaceaddexceptions{\csq@qclose@i}
\makeatletter

\usepackage{aliascnt}

\AtBeginDocument{ %

}

\title{Structural Results for High-Multiplicity Scheduling on Uniform Machines\thanks{This work was partially supported by DFG Project \href{https://gepris.dfg.de/gepris/projekt/335406402?context=projekt&task=showDetail&id=335406402&}{\enquote{Strukturaussagen und deren Anwendung in Scheduling- und Packungsprobleme}}, \texttt{JA 612 /20-1}}
}

\author{Hauke Brinkop\orcidID{0000-0002-7791-2353} \and
David Fischer\orcidID{0000-0001-8402-1818} \and
Klaus Jansen\orcidID{0000-0001-8358-6796}}
\authorrunning{H. Brinkop et al.}
\institute{Kiel University, Kiel, Germany \\ \email{\{hab,dfi,kj\}@informatik.uni-kiel.de}
}

\newcommand\es{\poly(\log n,\allowbreak \log m,\allowbreak \tau,\allowbreak d)}
\let\rel=\propto

\makeatletter
\def\cleartheorem#1{%
	\expandafter\let\csname#1\endcsname\relax
	\expandafter\let\csname c@#1\endcsname\relax
}
\makeatother	

\cleartheorem{theorem}
\spnewtheorem{theorem}{Theorem}{\bfseries}{\rmfamily}

\cleartheorem{definition}
\spnewtheorem{definition}[theorem]{Definition}{\bfseries}{\normalfont}

	\cleartheorem{lemma}
	\spnewtheorem{lemma}[theorem]{Lemma}{\bfseries}{\rmfamily}

	\cleartheorem{propo}
	\spnewtheorem{propo}[theorem]{Proposition}{\bfseries}{\rmfamily}

	\cleartheorem{corollary}
	\spnewtheorem{corollary}[theorem]{Corollary}{\bfseries}{\rmfamily}

	\cleartheorem{proposition}
	\spnewtheorem{proposition}[theorem]{Proposition}{\bfseries}{\rmfamily}

	\cleartheorem{definition}
	\spnewtheorem{definition}[theorem]{Definition}{\bfseries}{\rmfamily}
	
	\cleartheorem{observation}
	\spnewtheorem{observation}[theorem]{Observation}{\bfseries}{\rmfamily}

\newcommand\ie{i.e.\@\xspace}
\newcommand\eg{e.g.\@\xspace}

\begin{document}
	\maketitle

\begin{abstract}
Many optimization problems over integers are known to be NP-hard.
In order to achieve parameterized algorithms, often, \emph{proximity techniques} are used, i.e. exploiting that the solution to a fractional relaxation is not too far from an integral solution.
Parameterizing by the largest processing time $\pmax$ and the number of different job processing times $d$ (note that $d \leq \pmax$), we propose such a technique for High-Multiplicity Scheduling on Uniform Machines for the objectives Makespan Minimization (\Cmax) and Santa Claus (\Cmin) to obtain new structural results for these problems.
The novelty in our approach is that we deal with a fractional solution for only a sub-instance, where the sub-instance itself is not known a priori.
While the construction and computation of the fractional solution -- in contrast to usual proximity techniques -- is not done in polynomial time, this also allows us to formulate a comparably strong and general proximity statement.
Eventually, this allows us to reduce the number of jobs that need to be distributed to a polynomial in $\pmax$ for each machine and job type, by preassigning jobs according to the fractional solution, essentially returning a bounded number (at most $\O(\pmax^{\O(d^2)})$) of kernels, one for each (guessed) sub-instance.

We can use our structural results to obtain an algorithm with running time is $\pmax^{\mathcal O(d^2)} \cdot \es$, matching the best-known so far by \textcite{KNOP2021908}.

Moreover, we propose an $\pmax^{\mathcal O(d^2)} \cdot \es$ time algorithm for Envy Minimization \Cenvy in the High-Multiplicity Setting on Uniform Machines, showing that this problem is \textsc{fpt} in $\pmax$.

Eventually, we also propose a general mechanism to bound the largest coefficient in the Configuration ILP for so called \emph{Load Balancing Problems} by $(d \pmax)^{\O(d)}$, which we hope to be of interest for the development of algorithms.

\end{abstract}
\section{Introduction} \label{sec:introduction}
	In this paper, we consider \emph{High-Multiplicity Scheduling Problems On Uniform Machines}, where \enquote{high-multiplicity} refers to the following compact encoding:
	We are given $d \in \N$ job sizes in the form of a vector
	$p \in \N^d$ and a corresponding job multiplicity vector $\vecn \in \N_0^d$; and
	$\tau \in \N$ machine speeds in the form of a vector $s \in \N^\tau$ and a corresponding
	machine multiplicity vector $\vecm \in \N_0^\tau$. A job of size $p_j$ takes time
	$p_j / s_t$ to be processed on a machine of speed $s_t$.
	The task is to find an assignment of jobs to machines such that an objective function
	is optimized.

	Let $\super \sigma C_i$ denote the time when machine $i$ is finished
	processing the jobs assigned to it by a schedule $\sigma$.
	We consider the following
	three objectives: \Cmax, that is, find $\sigma$ such that $\max_i \super \sigma C_i$
	is minimal; \Cmin, that is, find $\sigma$ such that $\min_i \super \sigma C_i$
	is maximal; and \Cenvy, that is, find $\sigma$ such that
	$\max_i \super \sigma C_i - \min_i \super \sigma C_i$ is minimal.
	These problems are called $\QCmax[\HM]$, $\QCmin[\HM]$ and $\QCenvy[\HM]$ respectively in three-field notation~\cite{GRAHAM1979287}, where $\HM$ refers to the high-multiplicity encoding of both machines $m$ and jobs $n$.
	Note that for every pair of those objectives, there are instances where the set
	of optimal solutions is disjoint.	
	All three objectives (\Cmax, \Cmin, \Cenvy) find their applications in resource
	allocation.
	The objective \Cmax\ is one of the most classical scheduling objectives, if
	not the most classical one. It is also called \enquote{Makespan Minimization}
	as the expression $\max_i \super \sigma C_i$ is called the \emph{makespan}
	of $\sigma$. 
	The objective \Cmin\ finds applications in the sequencing
	of maintenance actions for modular
	gas turbine aircraft engines~\cite{DBLP:journals/orl/Woeginger97}.
	\textcite{DBLP:conf/stoc/BansalS06}
	called it the \enquote{Santa Claus Problem}.
	The objective \Cenvy{} is meant to measure some kind of fairness of an
	allocation, and was introduced by \textcite{DBLP:conf/sigecom/LiptonMMS04}.

	We are interested in the structure of optimal solutions if the largest job
	size $\pmax$ is small and there are only a few job types $d$.
	This occurs e.g. in approximative settings after rounding or when scheduling
	maintenance for air crafts~\cite{DBLP:journals/mor/McCormickSS01}.
	All of the above problems are NP-hard~\cite{DBLP:books/daglib/0030297} and hence unlikely to admit a
	polynomial time algorithm (unless $\textsc p = \textsc{np}$). However, there might be
	(and in fact, there are) polynomial time algorithms
	if $d$ and $\pmax$ are fixed; such algorithms are called \emph{slice-wise polynomial}
	(\textsc{xp}) with parameter $d + \pmax$.
	We can go even further for an algorithm with running time
	$\O(f(d,\pmax) \poly \abs I)$ for some function $f$, where $\abs I$ is the encoding length of the instance. Such algorithms
	are called \emph{fixed parameter-tractable (\textsc{fpt})} algorithms
	with parameter $d + \pmax$.
	Note that in contrast to low-multiplicity scheduling, where $\poly n$ is polynomial in the encoding of the input,
	for high-multiplicity $\poly n$ would be exponential in the encoding of the input.
	So, when designing \textsc{fpt} algorithms for the high-multiplicity setting, we have $\poly \abs I = \es$.
	
	\paragraph*{Related Work}
	Early works regarding high-multiplicity scheduling usually considered a
	single machine setting, e.\@{}g.\@{}~%
	\cite{DBLP:journals/dam/GranotS93,ToshioHamada2010KJ00006158866,DBLP:journals/ior/HochbaumS91,DBLP:journals/mp/HochbaumSS92,DBLP:journals/jacm/HochbaumS90,Posner85,DBLP:journals/ior/Psaraftis80}.
	This is inter alia due to the fact
	that for multiple machines, one cannot output the jobs assigned to a machine for
	each machine individually, as this would take $\Omega(m)$ space and hence $\Omega(m)$ time. 
	So, in order to handle multiple machines, a compact output encoding has to
	be used, usually together with some statements that show that there is always
	a solution that can be encoded in such a compact way. An approach of this kind
	was used by \textcite{DBLP:journals/mor/McCormickSS01}.
	\textcite{DBLP:journals/mp/CliffordP01} showed in \citeyear{DBLP:journals/mp/CliffordP01} that for objective \Cmax{} on uniform machines there
	is always a solution where only
	$\min \sing{m, \tau 2^d}$ different job assignments occur, which allows a sufficiently
	compact encoding for \textsc{fpt}-algorithms with any parameter $\kappa$ where
	$\kappa \geq d$. This argument can easily be extended to the objectives
	\Cmin{} and \Cenvy{}. Interestingly, the same argument was independently
	rediscovered in \citeyear{DBLP:journals/orl/EisenbrandS06} by
	\textcite{DBLP:journals/orl/EisenbrandS06}.
	They used it to show that 
	the so called \enquote{one dimensional Cutting Stock
	Problem}, a variation of bin packing with a high-multiplicity
	encoding of the items, is in \textsc{np}; this has been a long standing open question.
	\textcite{DBLP:conf/soda/GoemansR14} proved that the one dimensional cutting
	stock problem can be solved in polynomial time for a fixed number $d$
	of different item sizes, that is, the problem with parameter $d$ lies in \textsc{xp}.
	They also applied this result to high-multiplicity scheduling; in
	particular, they showed that with a compact encoding, Makespan Minimization on Uniform Machines
	with parameter $d + \tau$ lies in $\textsc{xp}$.
	\textcite{DBLP:conf/isaac/KouteckyZ20} showed that even when only the jobs are encoded in a compact way,
	Makespan Minimization on Uniform Machines is already \textsc{np}\hyp{}hard for 6 job types.
	Hence, there is no \textsc{fpt} algorithm with parameter $d$ for this problem unless $\textsc p = \textsc{np}$.

	\textcite{KNOP2021908} proposed a technique for solving $n$-fold ILPs, that is, a specific kind of block structured ILPs, where the blocks are encoded in a high-multiplicity way.
	Their approach is based on proximity of those ILPs.
	Applied to high-multiplicity scheduling, one achieves a running time of $\pmax^{\O(d^2)} \cdot \es$.
	\textcite{KK22} show that high-multiplicity scheduling on unrelated machines (and thus also on uniform machines) admits a kernel polynomial in $\pmax$ and the number of different machine types $\tau$, computable in time polynomial in these parameters.
	\textcite{mnich2013scheduling} introduced the idea of \enquote{balancing} an optimal solution
	around an optimal fractional solution if all machines are of the same speed, but with a rather large amount of jobs that is left to be scheduled per machine.
	
	\addcontentsline{toc}{subsection}{Related Work}%
	\label{sec:related_work}%

	\paragraph*{Our Contribution}
	Consider the classical dual approach in scheduling, i.e. guessing the optimum via binary search and solving the resulting feasibility problems using configurations.
	In \autoref{stmt:cutting} we show simple general mechanism to bound the largest coefficient in the Configuration ILP to $\pmax^{\O(d)}$ for so called \enquote{Load Balancing Problems} (c.f. \cite{DBLP:conf/icalp/BuchemRVW21}); i.e. scheduling problems where in the decision procedure mentioned above, the possible job assignments for a machine of type $t$ are all assignments whose processing time are between (machine-type-specific) given lower and upper bounds.

	If the machine capacities of the resulting decision instance are all bounded by a polynomial in $\pmax$, we can solve them in $\pmax^{\O(d^2)} \cdot \es$ using the ILP-algorithm by \textcite{DBLP:conf/soda/CslovjecsekEHRW21}.
	This motivates us to find a procedure to reduce machine capacities to a polynomial in $\pmax$.
	We consider three cases: In Case I, all machines (already) have a capacity $\leq d\pmax^2$.
	In Case II, the capacity of every machine exceeds $d\pmax^2$.
	In \autoref{stmt:lower_balancing_idle} we show that specific fractional solutions of an arbitrary selection of jobs, called fractional \emph{partial regular} schedules, can be used to preassign enough jobs such that each machine has remaining capacity bounded polynomial in $\pmax$.
	This process can be executed in polynomial time.

	Eventually, in Case III, we have both, machines of capacity $\leq d\pmax^2$ and $> d\pmax^2$.
	In \autoref{stmt:kernel} we show how to compute $\pmax^{\O(d^2)}$ many partial schedules in time $\pmax^{\O(d^2)} \es$ such that (i) the remainder capacities of all machines are bounded by a polynomial in $\pmax$ and (ii) at least on of them can be extended to a feasible schedule of all jobs.
	We show this using the results from \autoref{sec:balancing}, and by partitioning the jobs into those scheduled onto machines of capacity $\leq d\pmax^2$, and those scheduled on the remaining machines.
	While this result is no \enquote{true} kernelization (as opposed to \eg the result by \textcite{KK22}), as we need time exponential in the parameter to compute the reduced instance, and we get an exponential number of reduced instances, the size of this reduction is not depending on $\tau$.

	Moreover, our approach offers an interesting insight into the problem's difficulty, as there are two bottlenecks: Solving the ILP and partitioning the jobs.
	For the former, there has been some recent improvement achieving a running time of $\pmax^{\O(d)}$  \cite{JKPT24}.
	Thus, by speeding up the partitioning and using this recent result, one would directly get an algorithm for \QCmax and \QCmin faster than $\pmax^{\Theta(d^2)} \cdot \es$. 
	Interestingly, a similar behaviour can also be observed when approximating \QCmax~\cite{BBJMS23}.
	Note that our algorithm matches the current best running time for these problems of $\pmax^{\O(d^2)} \es$, achieved by the algorithm of \textcite{KNOP2021908}, but is of a more combinatorial type.

	We also develop a novel binary search routine for \Cenvy, where the naive approach did not lead to a (satisfying) result so far.
	Combining this routine with the result of \textcite{KNOP2021908} we derive a $\pmax^{\O(d^2)} \es$ time algorithm for objective \Cenvy.
	This, to the best of our knowledge, is the first result showing that \QCenvy is \textsc{fpt} only in the parameter $\pmax$, in the high-multiplicity setting or otherwise.

	\paragraph*{Organization of This Paper}
	In \autoref{sec:preliminaries} we start introducing basic terminology and notation.
	We introduce the general mechanism for bounding the largest coefficient for load balancing problems in \autoref{sec:general_load_balancing}.
	We then prove our proximity theorem in \autoref{sec:balancing} for instances with machines that are \enquote{large enough} ($\geq d\pmax^2$).
	Then, in \autoref{sec:fractional}, we show how we can bound $\smax T$ and $\tau$ in terms of $\pmax$ and $d$ for a given target value $T$, and extend our proximity result to general instances.
	Eventually, in \autoref{sec:algo}, we propose our algorithms for \Cmax, \Cmin and \Cenvy.
	Missing and full (formal) proofs can be found in the appendix.
	\section{Preliminaries and Notation} \label{sec:preliminaries}
	\providecommand\mysec[1]{%
		\vspace{0.4em}%
		\noindent\textbf{\large #1}.}
	
	\mysec{Numbers} For two numbers $a$ and $b$ we write
	$a \dotminus b$ for the positive difference of $a$ and $b$, that is, 
	$\max\sing{(a-b),0}$.
	For a set $M$ of positive integers, we write $\lcm M$
	for their lowest common multiple, the smallest positive integer that can be divided
	by every member of $M$.

	\mysec{Vectors}
	By convention, any lower case Latin letter with a subscript index is never a vector, but the component of a vector.
	For two vectors $v,w$ we write $v \leq w$ if for all $i$ we have
	$v_i \leq w_i$.
	We use $\floor v$ to denote the vector that one gets 
	by replacing every component $v_i$ of $v$ by $\floor {v_i}$.
	We use the same idea of point-wise rounding to define $\floor f$ for
	a map $f$. Analogously, $\ceil\cdot$ is defined for vectors and maps.
	For the special $d$\hyp{}dimensional vector that is zero in all entries,
	we write $\zerovec[d]$; analogously we define $\onevec[d]$ where all
	entries are one.
	We denote the support of a $d$\hyp{}dimensional
	vector $v$ by $\supp v \defeq \set{i \in \iv d | v_i \neq 0}$.
	Moreover, we extend the Bourbaki interval notation
	to arbitrary (partially ordered) sets and especially for vectors,
	e.\@{}g.\@{} write $\linterval{\zerovec[d]}{\onevec[d]}$
	for $\set{ v \in \R^d | \zerovec[d] \leq v \leq \onevec[d],\, v \neq \zerovec[d]}$.
	We use \enquote{$..$} as separator instead of \enquote{$;$} for intervals of integers
	and integer vectors.
	In the context of $\lcm(\cdot)$, $\max(\cdot)$, $\sum(\cdot)$, et cetera
	we use the same notation for vectors
	that is used for sets, e.\@{}g.\@{} write $\sum v$ for $\sum_{i \in \iv d} v_i$.

	\mysec{High-Multiplicity $\leftrightarrow$ Low-Multiplicity}
	As we are working in a high-multiplicity setting, machines do not have
	identities in the input. However, to show structural properties, we usually
	work and argue with single machines.
	To translate between the two settings,
	we assume there is a map $\resch \colon \iv m \to \iv \tau$ that maps
	each machine to its type. Note that we only use this map in proofs;
	in the algorithmic parts, we only work with machine types.

	\mysec{Scheduling}
	We call the tuple $I = (d,\tau,\vecn,\vecm, p, s)$ that contains all
	data of jobs and machines a \emph{scheduling instance} (cf.\@{} \autoref{sec:introduction}).
	We denote by $\imax \in \iv m$ an arbitrary yet fixed machine of maximal speed, that is, where $s_{\resch(i)} = \smax$. 
	A \emph{schedule} is a map of type
	$\iv \scalm \times \iv d \to \N_0$, mapping a machine $i \in \iv \scalm$
	and a job type $j \in \iv d$ to the (non-negative integer) number of jobs of type $j$ that are
	scheduled on machine $i$ -- note that a schedule has to distribute all the
	jobs of the instance and no more or less.
	Also note that this definition of a schedule
	refers to the mathematical entity, not the encoding.
	We write $\sigma_i(j)$ for the number of jobs of type $j$ that are assigned
	to machine $i$ by the schedule $\sigma$.
	As abbreviation we write $\sigma_i$ to denote
	the assignment vector $v$ of machine $i$, that is, $v_j = \sigma_i(j)$ for
	all $j \in \iv d$.
	By relaxing the requirement of schedules in the way that only a subset
	of the jobs of the instance might be distributed, we get a \emph{partial schedule}.
	Note that any schedule is also a partial schedule. We might also (not necessarily
	in combination with the aforementioned) relax the requirement that the assigned
	number of jobs has to be integer by allowing fractional values, too. This
	way we get what we call a \emph{fractional schedule}. By combining both
	relaxations, we get a  \emph{fractional partial schedule}.
	We also use the term \emph{integral schedule} to emphasize when we
	are not talking about a fractional schedule.  

	For a machine $i \in \iv m$ and a schedule $\sigma$
	we call $p^\intercal\sigma_i$ the \emph{load} of machine $i$ under $\sigma$,
	that is, the sum of the processing times of the 
	jobs assigned to $i$; and
	the time needed for the machine to process all the assigned load, that
	is load divided by the machine's speed,
	its \emph{completion time}.
	For a given completion time $T$ the \emph{capacity} of 
	machine $i \in \iv \scalm$ is given by
	$T \cdot s_{\resch(i)}$.
	We call the difference between a machine's capacity and its load
	the \emph{idle load} of the machine.
	Intuitively, the idle load of machine $i$ is the load
	that could be added (in the form of additional jobs) to machine $i$ while the completion time remains $\leq T$.
	When comparing schedules $\sigma$ and $\sigma'$, we call $\sigma$
	\enquote{smaller than or equal} to $\sigma'$ if $\sigma$ is point-wise
	smaller than or equal $\sigma'$. We call a schedule $\sigma$ an \emph{extension}
	of $\sigma'$ if $\sigma'$ is smaller than or equal to $\sigma$.
	An $n$-fold ILP is an ILP (Integer Linear Program) of
	size $r + st' \times nt'$ for $r,s,t',n \in \N$ consisting of
	$r$ arbitrary rows, and below them of a block diagonal matrix where
	each block has size $s \times t'$. Those ILPs can be solved
	in time $2^{\O(rs^2)}\enpar{rs\norm A_\infty}^{\O\enpar*{r^2s + s^2}}
	\enpar{nt'}^{1 + \mathrm o(1)}$ using the algorithm of
	\textcite{DBLP:conf/soda/CslovjecsekEHRW21}, which relies on ILP Proximity.
	By adding zero columns/rows, this result can also be applied to constraint
	matrices where the number of columns/rows is not equal over all blocks.
	A common approach in scheduling is to
	guess the optimal objective value $T$ via binary search, and then check if there
	exists a solution whose objective value is at least as good as $T$.
	This approach works 
	for a broad set of objectives and in particular for \Cmin, \Cmax, and \Cenvy.
	Therefore, often the so called
	\emph{Configuration ILP} is used. A \emph{configuration} is a vector
	$\in \N_0^d$ that encodes a feasible assignment of jobs to a specific
	machine type $t \in \iv \tau$. For the problems we consider, we can
	always express the set of configurations for each machine type $t \in \iv \tau$
	in the form $\mathcal C_t = \set{x \in \Interval{\zerovec[d]}{\bar \nu}  | \ell_t \leq p^\intercal x \leq u_t }$
	for some $\ell_t, u_t \in \Z \cup \sing{\infty, -\infty}$ and
	$\bar \nu \in \Interval{\zerovec[d]}{\nu}$.
	The configuration
	ILP now encodes that for each $t \in \iv \tau$, exactly $\mu_t$ configurations
	from the set $\mathcal C_t$ have to be chosen such that the total sum over
	all chosen	configurations is exactly $\nu$. For that, a variable $\super t x_c$
	is used for each $t \in \iv \tau$ and each $c \in \mathcal C_t$, indicating
	that configuration $c$ is chosen exactly $\super t x_c$ times for machines
	of type $t$.
	\begin{equation*}
	\tag{ConfILP} \label{ConfILP}
	\begin{aligned}
	\sum_{t \in \iv \tau} \sum_{c \in \mathcal C_t} {\super t x_c} c_j = \nu_j && \forall j \in \iv d \\
	\sum_{c \in \mathcal C_t} \super t x_c = \mu_t && \forall t \in \iv \tau
\\
	\super t x_c \in \N_0 && \forall t \in \iv \tau,\,c \in \mathcal C_t
	\end{aligned}
	\end{equation*}
	\ref{ConfILP} has $n$-fold structure with block parameters $r = d$, $t' = \max_{t \in \iv \tau} | \mathcal C_t |$, $s = 1$, $n = \tau$, and $\norm A_\infty$ equalling the largest entry in any $c \in \mathcal C_t$ over all $t \in \iv \tau$.
	The result of \textcite{KNOP2021908} implies that \ref{ConfILP} can be solved in time $\pmax^{\O(d^2)} \cdot \es$, choosing appropriate $\ell_t$ and $u_t$ for all $t \in \iv \tau$, and using very involved techniques.

	\section{A General Mechanism for Load Balancing Problems}
	\label{sec:general_load_balancing}
	The following lemma can be used to bound the largest coefficient in \ref{ConfILP} by $(d\pmax)^{\O(d)}$ for load balancing problems:
	\begin{lemma}[Cutting Lemma]
		\label{stmt:cutting}
		\normalfont
		For any $c \in \Interval{\zerovec[d]}{\vecn}$ where
		$p^\intercal c \geq
		d\pmax \lcm(p)$ there is $\bar c \in \Interval{\zerovec[d]}{c}$
		such that $p^\intercal \bar c = \lcm(p)$.
		\stopp
	\end{lemma}
	\begin{proof}
		Note that $d\pmax \norm c_\infty \geq \pmax \norm c_1 \geq p^\intercal c \geq d \pmax \lcm(p)$; hence also $\norm c_\infty \geq \lcm p$.
		By definition of $\norm \cdot _\infty$, there is $j \in \iv d$ such that $c_j \geq \lcm(p)$.
		Choose $\bar c$ by setting $\bar c_j \defeq \frac {\lcm (p)} {p_j}$ and for $k \in \iv d \setminus \sing j$,
		setting $c_k \defeq 0$. Observe that $\bar c$ is integral as  $p_j \mid \lcm(p)$.
		Eventually, observe that $p^\intercal \bar c = \frac {\lcm(p)} {p_j} p_j = \lcm(p)$.  
	\end{proof}
	Exhaustively applying this lemma achieves the desired result.
	More details can be found in \hyperref[sec:compr]{Appendix~\ref*{sec:compr}}.
	Moreover, scenarios such as Restricted Assignments (every job type $j \in \iv d$ may only be scheduled on a specific subset of machine types $\mathcal R_j \subseteq \iv \tau$) can be incorporated easily.

	\section{Proximity If Machines Are Large (Case II)} \label{sec:balancing}
	
	In this section, we  show a proximity bound for a specific kind of partial fractional schedules for machines of sufficiently large capacity. We then discuss how this can be exploited for objectives \Cmax and \Cmin. In \Autoref{sec:fractional} we discuss how this result can be used, even if machines of small capacity are also part of the instance (Case III).
	
	The idea of balancing in scheduling problems, which can be seen as a specific kind of proximity, is to exploit the possibly bounded difference (per machine and per job type)
	of a fractional schedule and an integral solution
	This idea was introduced by \textcite{mnich2013scheduling}.
	To get such a proximity, one may show that there always exists an optimal integral solution that is balanced similarly to the proportional fractional one.
	To do this, we want to make use of an exchange argument for load-invariant swapping of jobs between optimal integral solutions, in order to obtain a \enquote{more balanced} optimal integral solution.
	Such exchange arguments have already been previously used for e.g. knapsack~\cite{BHSS18} or coin change~\cite{chan_et_al:LIPIcs:2020:12895}.
	Since we need a slightly stronger version of the argument compared to the mentioned results, we use a version introduced by \textcite{Mnich}.
	As this result comes from an unpublished manuscript, we state a proof in \hyperref[sec:omitted]{Appendix~\ref*{sec:omitted}}.
	
	\begin{lemma}[Exchange Lemma \cite{BHSS18,chan_et_al:LIPIcs:2020:12895,Mnich}]
	\label{stmt:Mnich}
	\normalfont
		For any  $j \in \iv d$ and configuration vector
		$c \in \N_0^d$ satisfying $\norm c_1 \geq p_j$ there is a non-trivial configuration vector
		$\bar c \in \lInterval{\zerovec[d]}{c}$ such that
		$p^\intercal \bar c = \alpha p_j$ for some $\alpha \in \iv \pmax$.	\stopp
	\end{lemma}
	\noindent (Proof omitted, see \hyperref[sec:omitted]{Appendix~\ref*{sec:omitted}})

	For any machine $M_1$ that contains at least $\alpha \in \iv \pmax$ jobs of type $j$ in an optimal integral solution, \autoref{stmt:Mnich} can then be used to swap $\alpha$ jobs of type $j$ with some other jobs on a machine $M_2$ containing at least $p_j$ jobs of any types, without changing the load of any machine.
	For an example of such an exchange, see \autoref{fig:sched}.
	\begin{figure} \centering
			\begin{ganttchart} [hgrid, vgrid, x unit=3mm] 0 {21}
				\ganttgroup{
				}{8}{15} \\
				\newmachine[$i_1$] \addtasks {4 4 4 4 4} \\
				\newmachine[$i_2$] \addtasks {3 5 3 5}
				\ganttvrule{}{7} \ganttvrule{}{15}
			\end{ganttchart}
			\caption{Example of a possible exchange: We can swap two jobs with processing time $4$ from $i_1$ with one job of processing time $3$ and one of processing time $5$ from $i_2$.}
		\label{fig:sched}
	\end{figure}
	We now want to use this exchange argument to show proximity for a specific kind of fractional schedule, that always exists for instances with machines of sufficiently large capacities only.
	We first define what we mean by machines of sufficiently large capacity, which we simply call \emph{large}.
	\begin{definition}[Large Machines] \label{def:large_machines}
		\normalfont
		For some given completion time $T$, a machines $i$ is called large iff $s_{\resch(i)} T \geq d \pmax^2$. \stopp
	\end{definition}
	Machines that are not large are called \emph{small}.
	For our balancing lemma, the
	fractional schedule has to be of a specific kind, i.e. fulfil a property we refer to as \emph{regularity}:
	\begin{definition}[Regular Schedule] \label{def:regular}
		\normalfont
		A fractional partial schedule $\sigma$ is called \emph{regular} if for each
		job type $j \in \iv d$, either \lipicsitem{(i)}
		each machine gets at least $\pmax$ jobs
		of type $j$ or \lipicsitem{(ii)} each machine gets at most $\pmax$ jobs of type $j$. \stopp
	\end{definition}
	It is not too hard to compute such a schedule very efficiently for instances with large machines only,
	e.g.\@{} by putting $\min\sing{\frac{\nu_j}{m},\allowbreak \pmax}$ jobs of each job type $j \in \iv d$ on each machine and find an arbitrary fractional assignment of the remaining jobs
	(for $j \in \iv d$, if the minimum above is $\frac{\nu_j}m$, then (ii) of \autoref{def:regular} is satisfied, and in the other case (i) is satisfied).
	Thus, we can make the following observation.
	\begin{corollary}
	\label{cor:large_mach_reg}
	\normalfont
	If, for some given makespan bound $T$, all machines in a scheduling instance are large, there always exists a regular schedule for this instance.
	\stopp
	\end{corollary}
	Note that there are instances with both large and small machines where there is no regular fractional schedule at all: Consider an instance with $\pmax + 1$ jobs of processing time $\pmax$ and no other jobs types, with two machines $1$ and $2$ with speeds $s_1 = \pmax$ and $s_2 = 1$ respectively.
	Then any optimal solution will place $\pmax$ jobs on machine $1$ and $1$ job on machine $2$. In this example, $T = \pmax$, and thus here, machine $1$ is large and machine $2$ is small.
	We will deal with this issue later by splitting the instance into small and large machines.
	
	 We are now ready to show our proximity theorem for instances with large machines only.
	To simplify our theorem, we show it only for the feasibility problem \QCidle[\HM], which has the same inputs as \QCmax[\HM] together with an additional number $\kappa$, describing the maximum allowed idle load of any machine in an optimal solution.
	We give reductions from both \QCmin[\HM] and \QCmax[\HM] feasibility instances to feasibility instances of \QCidle[\HM] with $\kappa \in \Interval 0 \pmax$.
	The details of this reduction can be found in \hyperref[sec:conversion]{Appendix~\ref*{sec:conversion}}.
	Showing our result for this \QCidle[\HM] feasibility problem with idle load at most $\kappa$ then implies its applicability for both \QCmax[\HM] and \QCmin[\HM], which can then be applied to both optimization variants via standard binary search.
	Here, we only give a sketch of the proof. The full proof can be found in \hyperref[sec:omitted]{Appendix~\ref*{sec:omitted}}.

	\begin{restatable}[Proximity For Large Machines]{theorem}{stmtlowerbalancingidle}%
		\label{stmt:lower_balancing_idle}%
		\normalfont
		Let $\kappa \in \N_0$.
		Let $\hat\sigma$ be a fractional regular
		schedule with makespan $\leq T$.
		Then there is an integral schedule with makespan $\leq T$ where
		each machine $i \in \iv m$ has idle load $\leq \kappa$ 
		iff the following reduced fractional schedule
		\begin{align*}
			\check\sigma_i(j) \defeq
				\floor{\hat\sigma_i(j)} \dotminus
				\enpar*{\pmax + \floor*{\frac{\kappa}{\pmin}}}
				&& \forall i \in \iv m, j \in \iv d
		\end{align*}
		can be extended to a schedule with makespan $\leq T$ and idle load $\leq \kappa$ of the whole job set.
	\stopp\end{restatable}
	\input{proof___lower_balancing_idle___short.texb}
	We can now use the statement to preassign jobs according to $\check\sigma$.
	This allows us to reduce the capacities of all machines to a polynomial in $\pmax$ for both \QCmax[\HM] and \QCmin[\HM], essentially returning a kernel, for instances with large machines only.

	\section{Proximity in the general case (Case III)} \label{sec:fractional}
	So far, we have seen that if machines are of sufficiently large capacity, we can compute a regular fractional schedule easily and efficiently, allowing us to apply \autoref{stmt:lower_balancing_idle} in order to preassign jobs, and thus get a kernel where all machines have remaining capacities bounded in $\poly \pmax$.
	We have also seen that if there are machines that are not sufficiently large, there might not be a fractional regular schedule at all, which raises the question how this case can be handled, as \autoref{stmt:lower_balancing_idle} cannot be applied here.
	
	To do this, we want to solve the LP relaxation of the so-classed \emph{Assignment ILP}, an $n$-fold IP formulation for $\QCmax$ ($\QCmin$) where there is one assignment variable per job type and machine (not machine type), and the load of each machine of type $t \in \iv \tau$ is ensured to be $\leq s_t T$ ($\geq s_t T$) by a single constraint per machine.
	Note that classical polynomial time LP-algorithms have running times polynomial in $m$, the number of machines, which is not allowed in high-multiplicity scheduling.
	Instead, we use the recent result by \textcite[Proposition 21]{KJZ24}, which is a slight modification of the LP-algorithm from \textcite{DBLP:conf/soda/CslovjecsekEHRW21}, that allows to solve the LP-relaxation efficiently enough.
	We can then deduce by the proximity bound from \cite{DBLP:conf/soda/CslovjecsekEHRW21} that if the system is integer feasible, there is an integer solution not too far away from the computed fractional solution.
	Then the partition (which jobs are scheduled on large machines) of the jobs in that very integer solution is also not too far away from the partition of the jobs in the fractional one.
	We then enumerate all these possible job partitions and check feasibility by solving two independent problems, where one falls into Case I and the other into Case II.  
	A detailed proof can be found in \hyperref[sec:omitted]{Appendix~\ref*{sec:omitted}}.

	\begin{theorem} \label{stmt:kernel}
		\normalfont
		If an instance of \QCidle[\HM] is feasible with makespan $\leq T$ and idle load $\leq \kappa$, we can compute at most $\pmax^{\O(d^2)}$ partial schedules in time $\pmax^{\O(d^2)} \es$, such that
		\begin{enumerate}[(i)]
			\item  the remaining capacities of all machines are bounded by a polynomial in $\pmax$; and
			\item at least one of these partial schedules can be extended to an optimal solution by scheduling the remaining jobs on large machines only.
			\stopp
		\end{enumerate}
	\end{theorem}

	\section{Algorithmic Results} \label{sec:algo}
	In this section we describe some simple algorithmic applications we get based on our techniques.

	\subsection{Objectives \Cmax and \Cmin}
	Consider the following procedure:
	\begin{enumerate}
		\item Guess the optimal objective value $T$ via binary search.
		\item Use \autoref{stmt:kernel} to compute at most $\pmax^{\O(d^2)}$ many partial schedules with remaining capacities polynomial in $\pmax$.
		\item Set up \ref{ConfILP} with capacities, and thus the largest ILP coefficient, bounded polynomial in $\pmax$, for each of the partial schedules.
		\item Solve this \ref{ConfILP} in time $\pmax^{\O(d^2)} \cdot \es$ using the $n$-fold ILP algorithm by \textcite{DBLP:conf/soda/CslovjecsekEHRW21}.
	\end{enumerate}
	This returns an optimal solution for instances of both $\QCmax[\HM]$ and $\QCmin[\HM]$ in time $\pmax^{\O(d^2)} \cdot \es$, matching the best known running time by \textcite{KNOP2021908} for these problems.

	\subsection{Objective \Cenvy}

	The main difficulty for \Cenvy compared to \Cmax and \Cmin is that, given some $k$, it is not that simple to decide if $k \leq \OPT$.
	This is mainly due to the fact that in order to use a configuration ILP, we do need lower and upper bounds for the completion times.
	But finding these is not that simple, as from infeasibility of an instance with lower bound $T - k$ and upper bound $T$ for some guess $T$, it is not clear if we need to increase or decrease $T$, in contrast to \Cmax and \Cmin, where this is very clear.
	Many approaches to tackle this have running time $\Omega(\smax)$, which is not \textsc{fpt} with respect to $\pmax$.
	We propose the first approach to solve this problem without the need of introducing an additional parameter\footnote{For example, if one does binary search for $\OPT$ and introduces a variable $T$ denoting the minimum completion time, \ie all completion times have to be in $\interval{T}{T + \OPT}$, then the largest coefficient does depend on $\smax$. Due to this, ILP results like \cite{lin4block} do not yield the desired result (as we discussed with the authors).}, which eventually leads to the first \textsc{fpt} algorithm with parameter $\pmax$ for \QCenvy.
	\[
		\Cmax(\sigma) \defeq \max_{i \in \iv m} s_i^\inv p^\intercal \sigma_i
		\qquad \text{and} \qquad 
		\Cmin(\sigma) \defeq \min_{i \in \iv m} s_i^\inv p^\intercal \sigma_i
		\enspace .
	\]
	Then the objective value $\Cenvy = \Cmax(\sigma) - \Cmin(\sigma)$ for schedule $\sigma$. Moreover, a machine $i$ is \emph{extremal} under $\sigma$ if $s_i^\inv p^\intercal \sigma_i \in \set{\Cmin(\sigma), \Cmax(\sigma)}$.
	Let
$
		A \defeq \frac{p^\intercal \nu}{s^\intercal \mu}.
$
	A machine $i$ is \emph{nice} if $p^\intercal \sigma_i \in \interval {s_i A - \pmax} {s_i A + \pmax}$. First not the following simple property.
	\begin{proposition}
	\label{prop:cmaxacmin}
	\normalfont
		For any schedule $\sigma$ we either have $\Cmax(\sigma) > A > \Cmin(\sigma)$ or $\Cmax(\sigma) = A = \Cmin(\sigma)$.
	\stopp
	\end{proposition}
	\noindent (Proof omitted, see \hyperref[sec:omitted]{Appendix~\ref*{sec:omitted}})

We can also show that an optimal schedule for $\Cenvy$ always contains a machine that is both extremal and nice.

	\begin{lemma}
		\label{stmt:exists_extremal}
		\normalfont
		For $m > 1$, there is a schedule $\sigma$ minimizing envy, i.e. minimizing $\Cmax(\sigma) - \Cmin(\sigma)$, such that an extremal machine is nice.
		\stopp
	\end{lemma} 
	\begin{proof}
		Let $\sigma$ be an optimal solution that minimizes the number of extremal machines.
		We proceed with proof by contradiction, assuming there is no extremal machine that is nice.
		Note that there have to be at least two extremal machines.
		Let $i_1$ such that $\Cmax(\sigma) = s_{i_1}^\inv p^\intercal \sigma_{i_1}$ and $i_2 \neq i_1$ such that $\Cmin(\sigma) = s_{i_2}^\inv p^\intercal \sigma_{i_2}$.
		By assumption, neither of those two is nice, i.e.
		\[
			p^\intercal \sigma_{i_1} > s_{i_1} A + \pmax
			\qquad
			\text{and}
			\qquad
			p^\intercal \sigma_{i_2} < s_{i_2} A - \pmax
			\enspace .
		\]
		We can thus move an arbitrary job from $i_1$ to $i_2$ without increasing $\Cmax(\sigma)$ or decreasing $\Cmin(\sigma)$.
		Thus, the solution does not become sub-optimal.
		By doing so, either the number of extremal machines is reduced, which contradicts that $\sigma$ has a minimal number of extremal machines; or either $i_1$ or $i_2$ remain extremal, which contradicts the optimality of $\sigma$.
		\qed
	\end{proof}
	We can exploit this algorithmically by basically guessing the type of an extremal machine, guessing its load, which allows us to derive either the lower or upper bound on the completion times; and then finding the complementary bound via binary search.
	\begin{theorem}
	\label{stmt:cenvy_fpt_pmax}
	\normalfont
		There is an algorithm solving \QCenvy[\HM] in time \\ $\pmax^{\O(d^2)}\allowbreak \cdot \es$.
	\stopp
	\end{theorem}
	\noindent (Proof omitted, see \hyperref[sec:omitted]{Appendix~\ref*{sec:omitted}})
	
	As $d \leq \pmax$, this implies that \QCenvy[\HM] in \textsc{fpt} with parameter $\pmax$, which, to the best of our knowledge, was not known before. 
	
	\section{Conclusion}
	We introduced two techniques: A general mechanism for reducing the maximum machine speed
	for general load balancing problems such that afterwards it can be bounded by $\O((d\pmax)^{\O(d)})$,
	which might be of use for other problems by itself (it can be applied directly to e.g. Restricted Assignment);
	and a sophisticated proximity statement that gave us novel insights
	into the structure of optimal solutions.
	It turns out that the hard part is to decide
	which jobs are to be scheduled on slow machines -- a behaviour that can also be observed in the approximate setting~\cite{BBJMS23}:
	In the currently fastest EPTAS, if there are only large machines (using a different threshold), one can replace the MILP solver by a faster ILP solver.
	On the other hand, if there are only small machines, one can simply use an LP-solver.
	Only if both kinds of machines are present, the MILP Algorithm is needed.

	Given the partition of the jobs, every (regular)
	fractional partial schedule with appropriate makespan can be extended to an integral
	schedule on the faster machines after priorly removing just a few ($\poly \pmax$) jobs
	per machine and job type.
	We used this result to essentially return $\O(\pmax^{\O(d^2)})$ kernels per instance of $\QCmax[\HM]$ and $\QCmin[\HM]$, where there are at most $\poly \pmax$ unassigned jobs per machine and job type.
	We imagine that our theorems find application in further work,
	\eg in scheduling with different objective functions, as basis for other structural
	theorems or in other combinatorial problems.
	
	Based on our structural theorems we then presented algorithms for both \Cmax and \Cmin, matching the currently best running time. It would be interesting to see
	if the quadratic dependency on $d$ in the exponent is necessary; so far, better bounds
	are only known for identical machines, where algorithms only have linear dependency on $d$ in the
	exponent.
	A recent result by \textcite[Theorem 2]{JKPT24} showed that the \ref{ConfILP} can be solved in time $(\pmax \tau)^{\O(d)} \log(m)$ if the largest coefficient is bounded by a polynomial in $\pmax$.
	So, an efficient partition procedure could not only help to speed up the EPTAS, but would also directly lead to an algorithm for $\threefield Q {\HM} {\sing{\Cmax, \Cmin}}$ with running time $\pmax^{o(d^2)} \cdot \es$.
	We thus believe this to be an interesting direction for future research.
	Moreover, we showed that the problem \Cenvy is \textsc{fpt} in $\pmax$ alone by combining our structural results with a novel binary search routine on the objective value of this problem.

	\FloatBarrier%
	\printbibliography

@inproceedings{KK22,
  author       = {Dusan Knop and
                  Martin Kouteck{\'{y}}},
  editor       = {Shiri Chechik and
                  Gonzalo Navarro and
                  Eva Rotenberg and
                  Grzegorz Herman},
  title        = {Scheduling Kernels via Configuration {LP}},
  booktitle    = {30th Annual European Symposium on Algorithms, {ESA} 2022, September
                  5-9, 2022, Berlin/Potsdam, Germany},
  series       = {LIPIcs},
  volume       = {244},
  pages        = {73:1--73:15},
  publisher    = {Schloss Dagstuhl - Leibniz-Zentrum f{\"{u}}r Informatik},
  year         = {2022},
  url          = {https://doi.org/10.4230/LIPIcs.ESA.2022.73},
  doi          = {10.4230/LIPICS.ESA.2022.73},
  bibsource    = {dblp computer science bibliography, https://dblp.org}
}

@article{DBLP:journals/orl/EisenbrandS06,
  author    = {Friedrich Eisenbrand and
               Gennady Shmonin},
  title     = {Carath{\'{e}}odory bounds for integer cones},
  journal   = {Operations Research Letters.},
  volume    = {34},
  number    = {5},
  pages     = {564--568},
  year      = {2006},
  url       = {https://doi.org/10.1016/j.orl.2005.09.008},
  doi       = {10.1016/j.orl.2005.09.008},
  bibsource = {dblp computer science bibliography, https://dblp.org}
}

@misc{Mnich,
  author = {Matthias Mnich and Alisa Govzmann and Simon Omlor},
  title = {Faster Algorithms for Parallel and Related Machine Scheduling},
  year = {2023},
  note = {Unpublished Manuscript}
}

@article{DBLP:journals/mp/CliffordP01,
  author    = {John J. Clifford and
               Marc E. Posner},
  title     = {Parallel machine scheduling with high multiplicity},
  journal   = {Mathematical Programming},
  volume    = {89},
  number    = {3},
  pages     = {359--383},
  year      = {2001},
  url       = {https://doi.org/10.1007/PL00011403},
  doi       = {10.1007/PL00011403},
  bibsource = {dblp computer science bibliography, https://dblp.org}
}

@inproceedings{DBLP:conf/isaac/KouteckyZ20,
  author    = {Martin Kouteck{\'{y}} and
               Johannes Zink},
  editor    = {Yixin Cao and
               Siu{-}Wing Cheng and
               Minming Li},
  title     = {Complexity of Scheduling Few Types of Jobs on Related and Unrelated
               Machines},
  booktitle = {31st International Symposium on Algorithms and Computation, {ISAAC}
               2020, December 14-18, 2020, Hong Kong, China (Virtual Conference)},
  series    = {LIPIcs},
  volume    = {181},
  pages     = {18:1--18:17},
  publisher = {Schloss Dagstuhl -- Leibniz-Zentrum f{\"{u}}r Informatik},
  year      = {2020},
  url       = {https://doi.org/10.4230/LIPIcs.ISAAC.2020.18},
  doi       = {10.4230/LIPIcs.ISAAC.2020.18},
  bibsource = {dblp computer science bibliography, https://dblp.org}
}

@inproceedings{DBLP:conf/soda/GoemansR14,
  author    = {Michel X. Goemans and
               Thomas Rothvo{\ss}},
  editor    = {Chandra Chekuri},
  title     = {Polynomiality for Bin Packing with a Constant Number of Item Types},
  booktitle = {Proceedings of the Twenty-Fifth Annual {ACM-SIAM} Symposium on Discrete
               Algorithms, {SODA} 2014, Portland, Oregon, USA, January 5--7, 2014},
  pages     = {830--839},
  publisher = {{SIAM}},
  year      = {2014},
  url       = {https://doi.org/10.1137/1.9781611973402.61},
  doi       = {10.1137/1.9781611973402.61},
  bibsource = {dblp computer science bibliography, https://dblp.org}
}

@inproceedings{DBLP:conf/soda/CslovjecsekEHRW21,
  author    = {Jana Cslovjecsek and
               Friedrich Eisenbrand and
               Christoph Hunkenschr{\"{o}}der and
               Lars Rohwedder and
               Robert Weismantel},
  editor    = {D{\'{a}}niel Marx},
  title     = {Block-Structured Integer and Linear Programming in Strongly Polynomial
               and Near Linear Time},
  booktitle = {Proceedings of the 2021 {ACM-SIAM} Symposium on Discrete Algorithms,
               {SODA} 2021, Virtual Conference, January 10--13, 2021},
  pages     = {1666--1681},
  publisher = {{SIAM}},
  year      = {2021},
  url       = {https://doi.org/10.1137/1.9781611976465.101},
  doi       = {10.1137/1.9781611976465.101},
  bibsource = {dblp computer science bibliography, https://dblp.org}
}

@inproceedings{DBLP:conf/icalp/BuchemRVW21,
  author    = {Moritz Buchem and
               Lars Rohwedder and
               Tjark Vredeveld and
               Andreas Wiese},
  editor    = {Nikhil Bansal and
               Emanuela Merelli and
               James Worrell},
  title     = {Additive Approximation Schemes for Load Balancing Problems},
  booktitle = {48th International Colloquium on Automata, Languages, and Programming,
               {ICALP} 2021, July 12--16, 2021, Glasgow, Scotland (Virtual Conference)},
  series    = {LIPIcs},
  volume    = {198},
  pages     = {42:1--42:17},
  publisher = {Schloss Dagstuhl -- Leibniz-Zentrum f{\"{u}}r Informatik},
  year      = {2021},
  url       = {https://doi.org/10.4230/LIPIcs.ICALP.2021.42},
  doi       = {10.4230/LIPIcs.ICALP.2021.42},
  bibsource = {dblp computer science bibliography, https://dblp.org}
}

@book{DBLP:books/daglib/0030297,
  author    = {David P. Williamson and
               David B. Shmoys},
  title     = {The Design of Approximation Algorithms},
  publisher = {Cambridge University Press},
  year      = {2011},
  url       = {http://www.cambridge.org/de/knowledge/isbn/item5759340/?site\_locale=de\_DE},
  isbn      = {978-0-521-19527-0},
  bibsource = {dblp computer science bibliography, https://dblp.org}
}

@inproceedings{DBLP:conf/stoc/BansalS06,
  author    = {Nikhil Bansal and
               Maxim Sviridenko},
  editor    = {Jon M. Kleinberg},
  title     = {The Santa Claus problem},
  booktitle = {Proceedings of the 38th Annual {ACM} Symposium on Theory of Computing, {STOC} 2006, May 21--23, 2006,
               Seattle, USA},
  pages     = {31--40},
  publisher = {{ACM}},
  year      = {2006},
  url       = {https://doi.org/10.1145/1132516.1132522},
  doi       = {10.1145/1132516.1132522},
  bibsource = {dblp computer science bibliography, https://dblp.org}
}

@article{DBLP:journals/orl/Woeginger97,
  author    = {Gerhard J. Woeginger},
  title     = {A polynomial-time approximation scheme for maximizing the minimum
               machine completion time},
  journal   = {Operations Research Letters.},
  volume    = {20},
  number    = {4},
  pages     = {149--154},
  year      = {1997},
  url       = {https://doi.org/10.1016/S0167-6377(96)00055-7},
  doi       = {10.1016/S0167-6377(96)00055-7},
  bibsource = {dblp computer science bibliography, https://dblp.org}
}

@article{GRAHAM1979287,
	title = "Optimization and Approximation in Deterministic Sequencing and Scheduling: A Survey",
	author = "Graham, Ronald L. and Lawler, Eugene L. and Lenstra, Jan K. and Rinnooy Kan, Alexander H. G.",
	year = "1979",
	doi = "10.1016/S0167-5060(08)70356-X",
	language = "English",
	volume = "5",
	pages = "287--326",
	journal = "Annals of Discrete Mathematics",
	issn = "0167-5060",
	publisher = "Elsevier"
}

@inproceedings{DBLP:conf/sigecom/LiptonMMS04,
  author    = {Richard J. Lipton and
               Evangelos Markakis and
               Elchanan Mossel and
               Amin Saberi},
  editor    = {Jack S. Breese and
               Joan Feigenbaum and
               Margo I. Seltzer},
  title     = {On approximately fair allocations of indivisible goods},
  booktitle = {Proceedings 5th {ACM} Conference on Electronic Commerce (EC-2004),
               New York, USA, May 17--20, 2004},
  pages     = {125--131},
  publisher = {{ACM}},
  year      = {2004},
  url       = {https://doi.org/10.1145/988772.988792},
  doi       = {10.1145/988772.988792},
  bibsource = {dblp computer science bibliography, https://dblp.org}
}

@article{DBLP:journals/ior/Psaraftis80,
  author    = {Harilaos N. Psaraftis},
  title     = {A Dynamic Programming Approach for Sequencing Groups of Identical
               Jobs},
  journal   = {Operations Research},
  volume    = {28},
  number    = {6},
  pages     = {1347--1359},
  year      = {1980},
  url       = {https://doi.org/10.1287/opre.28.6.1347},
  doi       = {10.1287/opre.28.6.1347},
  bibsource = {dblp computer science bibliography, https://dblp.org}
}

@misc{Posner85,
  author = {Marc E. Posner},
  year = {1985},
  title = {The Complexity of Earliness and Tardiness Scheduling Problems under Id-Encoding},
  type = {Working Paper 85--70},
  institution = {New York University}
}

@article{DBLP:journals/ior/HochbaumS91,
  author    = {Dorit S. Hochbaum and
               Ron Shamir},
  title     = {Strongly Polynomial Algorithms for the High Multiplicity Scheduling
               Problem},
  journal   = {Operations Research},
  volume    = {39},
  number    = {4},
  pages     = {648--653},
  year      = {1991},
  url       = {https://doi.org/10.1287/opre.39.4.648},
  doi       = {10.1287/opre.39.4.648},
  bibsource = {dblp computer science bibliography, https://dblp.org}
}

@article{DBLP:journals/mp/HochbaumSS92,
  author    = {Dorit S. Hochbaum and
               Ron Shamir and
               J. George Shanthikumar},
  title     = {A polynomial algorithm for an integer quadratic non-separable transportation
               problem},
  journal   = {Mathematical Programming},
  volume    = {55},
  pages     = {359--371},
  year      = {1992},
  url       = {https://doi.org/10.1007/BF01581207},
  doi       = {10.1007/BF01581207},
  bibsource = {dblp computer science bibliography, https://dblp.org}
}

@article{DBLP:journals/jacm/HochbaumS90,
  author    = {Dorit S. Hochbaum and
               J. George Shanthikumar},
  title     = {Convex Separable Optimization Is Not Much Harder than Linear Optimization},
  journal   = {Journal of the {ACM}},
  volume    = {37},
  number    = {4},
  pages     = {843--862},
  year      = {1990},
  url       = {https://doi.org/10.1145/96559.96597},
  doi       = {10.1145/96559.96597},
  bibsource = {dblp computer science bibliography, https://dblp.org}
}

@article{ToshioHamada2010KJ00006158866,
  title={A Bayesian Sequential Single Machine Batching and Scheduling Problem with Random Setup Time},
  author={Toshio Hamada},
  journal={Journal of the Operations Research Society of Japan},
  volume={53},
  number={1},
  pages={79--100},
  year={2010},
  doi={10.15807/jorsj.53.79}
}

@article{DBLP:journals/dam/GranotS93,
  author    = {Frieda Granot and
               Jadranka Skorin{-}Kapov},
  title     = {On Polynomial Solvability of the High Multiplicity Total Weighted
               Tardiness Problem},
  journal   = {Discrete Applied Mathematics},
  volume    = {41},
  number    = {2},
  pages     = {139--146},
  year      = {1993},
  url       = {https://doi.org/10.1016/0166-218X(93)90034-L},
  doi       = {10.1016/0166-218X(93)90034-L},
  bibsource = {dblp computer science bibliography, https://dblp.org}
}

@article{DBLP:journals/mor/McCormickSS01,
  author    = {S. Thomas McCormick and
               Scott R. Smallwood and
               Frits C. R. Spieksma},
  title     = {A Polynomial Algorithm for Multiprocessor Scheduling with Two Job
               Lengths},
  journal   = {Mathematics of Operations Research},
  volume    = {26},
  number    = {1},
  pages     = {31--49},
  year      = {2001},
  url       = {https://doi.org/10.1287/moor.26.1.31.10590},
  doi       = {10.1287/moor.26.1.31.10590},
  bibsource = {dblp computer science bibliography, https://dblp.org}
}

@misc{mnich2013scheduling,
      title={Scheduling Meets Fixed-Parameter Tractability}, 
      author={Matthias Mnich and Andreas Wiese},
      year={2013},
      eprint={1311.4021},
      archivePrefix={arXiv},
      primaryClass={cs.DS}
}

@misc{KJZ24,
      title={Exact and Approximate High-Multiplicity Scheduling on Identical Machines}, 
      author={Klaus Jansen and Kai Kahler and Esther Zwanger},
      year={2024},
      eprint={2404.17274},
      archivePrefix={arXiv},
      primaryClass={cs.DS},
      url={https://arxiv.org/abs/2404.17274}, 
}

@article{KNOP2021908,
title = {Parameterized complexity of configuration integer programs},
journal = {Operations Research Letters},
volume = {49},
number = {6},
pages = {908-913},
year = {2021},
issn = {0167-6377},
doi = {https://doi.org/10.1016/j.orl.2021.11.005},
url = {https://www.sciencedirect.com/science/article/pii/S0167637721001565},
author = {Dušan Knop and Martin Koutecký and Asaf Levin and Matthias Mnich and Shmuel Onn}
}

@InProceedings{chan_et_al:LIPIcs:2020:12895,
  author =  {Timothy M. Chan and Qizheng He},
  title = {{More on Change-Making and Related Problems}},
  booktitle = {28th Annual European Symposium on Algorithms (ESA 2020)},
  pages = {29:1--29:14},
  series =  {Leibniz International Proceedings in Informatics (LIPIcs)},
  ISBN =  {978-3-95977-162-7},
  ISSN =  {1868-8969},
  year =  {2020},
  volume =  {173},
  editor =  {Fabrizio Grandoni and Grzegorz Herman and Peter Sanders},
  publisher = {Schloss Dagstuhl--Leibniz-Zentrum f{\"u}r Informatik},
  address = {Dagstuhl, Germany},
  URL =   {https://drops.dagstuhl.de/opus/volltexte/2020/12895},
  URN =   {urn:nbn:de:0030-drops-128958},
  doi =   {10.4230/LIPIcs.ESA.2020.29}
}

@inproceedings{BHSS18,
author = {Bateni, MohammadHossein and Hajiaghayi, MohammadTaghi and Seddighin, Saeed and Stein, Cliff},
title = {Fast algorithms for knapsack via convolution and prediction},
year = {2018},
isbn = {9781450355599},
publisher = {Association for Computing Machinery},
address = {New York, NY, USA},
url = {https://doi.org/10.1145/3188745.3188876},
doi = {10.1145/3188745.3188876},
booktitle = {Proceedings of the 50th Annual ACM SIGACT Symposium on Theory of Computing},
pages = {1269–1282},
numpages = {14},
location = {Los Angeles, CA, USA},
series = {STOC 2018}
}

@misc{JKPT24,
      title={Improving the Parameter Dependency for High-Multiplicity Scheduling on Uniform Machines}, 
      author={Klaus Jansen and Kai Kahler and Lis Pirotton and Malte Tutas},
      year={2024},
      eprint={2409.04212},
      archivePrefix={arXiv},
      primaryClass={cs.DS},
      url={https://arxiv.org/abs/2409.04212}, 
}

@article{lin4block,
	author    = {Chen, Hua and Chen, Lin and Zhang, Guochuan},
	title     = {FPT algorithms for a special block-structured integer program with applications in scheduling},
	journal   = {Mathematical Programming},
	year      = {2024},
	url       = {https://doi.org/10.1007/s10107-023-02046-z},
	doi       = {10.1007/s10107-023-02046-z},
}

@inproceedings{BBJMS23,
  author       = {Sebastian Berndt and
                  Hauke Brinkop and
                  Klaus Jansen and
                  Matthias Mnich and
                  Tobias Stamm},
  editor       = {Satoru Iwata and
                  Naonori Kakimura},
  title        = {New Support Size Bounds for Integer Programming, Applied to Makespan
                  Minimization on Uniformly Related Machines},
  booktitle    = {34th International Symposium on Algorithms and Computation, {ISAAC}
                  2023, December 3-6, 2023, Kyoto, Japan},
  series       = {LIPIcs},
  volume       = {283},
  pages        = {13:1--13:18},
  publisher    = {Schloss Dagstuhl - Leibniz-Zentrum f{\"{u}}r Informatik},
  year         = {2023},
  url          = {https://doi.org/10.4230/LIPIcs.ISAAC.2023.13},
  doi          = {10.4230/LIPICS.ISAAC.2023.13},
}
	\clearpage
	\appendix
	\addcontentsline{toc}{section}{Appendix}
	\renewcommand\appproof[1]{Proof of
		\autoref{#1} (page \pageref*{#1})%
	}

	\clearpage
	\section{Omitted Proofs}
	\label{sec:omitted}
	
	\newcommand\pof[1]{\underline{Proof of \autoref{#1}:}}
	\pof{stmt:Mnich}
	\begin{proof}
		We may write $c = \sum_{i \in \iv{\norm c_1}} \super i e$
		for some unit vectors $\super i e$, i.e. $\super i e \in \set { z \in \sing{0,1}^d | \norm z_1 = 1 }$ for all $i$.
		Now consider the sequence $(a_i)_{i \in \iv {p_j}}$
		where
		\[ a_i \defeq p^\intercal \enpar*{\sum_{k \in \iv i} \super k e} \mod p_j. \]
		Note that for all $i$ we have $0 \leq a_i \leq p_j-1$.
		Moreover, $(a_i)_{i \in \iv {p_j}}$ has $p_j$ elements. If there is $1 \leq i \leq p_j$ such that $a_i = 0$, then
		$\bar c \defeq \sum_{k \in \iv i} \super k e$ has the desired property.
		Otherwise, by pigeonhole, there are $1 \leq i < i' \leq p_j$ such that	$a_i = a_{i'}$.
		Then
		$\bar c \defeq \sum_{k \in \iv {i'}} \super k e - \sum_{k \in \iv i} \super k e = \sum_{k \in \iv{i'} \setminus \iv i} \super k e$ has the desired property.
		\qed
	\end{proof}
	\pof{stmt:lower_balancing_idle}
	\input{proof___lower_balancing_idle___long.texb}
	\pof{stmt:kernel}
	\input{proof___balancing_idle_all_machines___long.texb}	
	\pof{prop:cmaxacmin}
	
		\begin{proof}
		Observe that
		$
			\sum_{i \in \iv m} s_i A = A \sum_{i \in \iv m} s_i = \frac{p^\intercal \nu}{s^\intercal \mu} \sum_{i \in \iv m} s_i = p^\intercal \nu.
		$
		We thus have
		$
			\sum_{i \in \iv m} p^\intercal \sigma_i
			= \sum_{i \in \iv m} s_i A
			\iff\sum_{i \in \iv m} (s_i^\inv  p^\intercal \sigma_i - A) = 0.
		$
		If one of the summands is non-zero, there has to be another non-zero summand of opposite sign.
		This shows the claim.
		\qed
	\end{proof}	
	\pof{stmt:cenvy_fpt_pmax}
	\begin{proof}
	We proceed as follows.
	Remember that the set of configurations $\mathcal C_t$ is defined as $\mathcal C_t = \set{x \in \Interval{\zerovec[d]}{\bar \nu}  | \ell_t \leq p^\intercal x \leq u_t }$.
		\begin{enumerate}
			\item Try the case $\Cmax(\sigma) = \Cmin(\sigma)$ by simply setting $\ell_t \gets s_t A$, $u_t \gets s_t A$, setting up \ref{ConfILP}
			\item Try to solve \ref{ConfILP} using the algorithm of \textcite{KNOP2021908}.
			\item If successful, we have $\OPT = 0$. Return the computed schedule
			\item If unsuccessful, then $\OPT > 0$.
			\item Guess the type $t$ of a nice machine that is extremal, whose existence is ensured by \autoref{stmt:exists_extremal}.
			\item Guess the load of a nice extremal machine of type $t$, which is between $s_t A - \pmax$ and $s_t A + \pmax$ (at most $\O(\pmax)$ possible values). As we have $\Cmax(\sigma) > A > \Cmin(\sigma)$, we thus know either $\Cmax(\sigma)$ or $\Cmin(\sigma)$.			\item  Find the corresponding $\Cmin(\sigma)$ or $\Cmax(\sigma)$ value via binary search.
			\item Set $\ell_t \defeq \ceil*{\Cmin(\sigma)s_t}$ and $u_t \defeq \floor*{\Cmax(\sigma)s_t}$, set up \ref{ConfILP}
			\item Solve \ref{ConfILP} using the algorithm of \textcite{KNOP2021908}.
			\item For all feasible choices of $\ell$ and $u$, find the pair of bounds that has the best objective value and return the corresponding schedule computed priorly.
	\end{enumerate}
	
	As the guessing of a nice extremal machine of type $t$, and guessing its load takes at most time $\O(\tau \pmax)$, the running time is dominated by the time needed to solve \ref{ConfILP}. As we have only $\O(\tau)$ different block types in \ref{ConfILP} this yields an overall running time of $\pmax^{\O(d^2)} \cdot \es$.
	\end{proof}
	
	\section{Compression}
	\label{sec:compr}
	In this section, we give a formal and detailed proof for the reduction of the maximum capacity to $\pmax^{\O(d)}$, a process we call \emph{compression}.

	\providecommand\confs{\mathcal C}
	First, let us formally consider on how to reduce the \enquote{capacity} of the machines.
	More precisely: Consider a guess $\alpha$ for the optimum objective value.
	For both, \Cmax and \Cmin, we construct an instance of our \Cmax variation, where the idle load
	has to be $\leq \kappa$.
	Let $T$ be the (corresponding, stems from the construction mechanism) upper bound on the makespan.
	Then, in \ref{ConfILP}, for each machine $i \in \iv m$ we want to select a configuration $c$ whose load lies within $\interval{s_{\resch(i)} (T \dotminus \kappa)}{s_\resch(i) T}$.

	Now, if $Ts_{\resch(i)}$, i.e. the machine's capacity in the constructed instance, is too large we can split the machine into two machines of smaller capacity using \autoref{stmt:cutting} (it depends on the objective considered, which sub-statement is to be used).
	For that, we first need the following statement, which shows that the combinations (via element-wise addition) of the configuration sets of the two resulting machines are yield exactly the original set of configurations.
	For readability purposes, let $\mydelta := \lcm(p) \leq \pmax^d$ and $\mygamma := d \pmax \delta$.
	\begin{lemma} \label{stmt:red}
		\normalfont
 		Let $\confs(u,\ell) \defeq \set{c \in \N_0^d | \ell \leq p^\intercal c \leq u}$.
		For any $\ell \in \N_0$ and $u \in \N_0 \uplus \sing{\infty}$ the following statements hold:
		\begin{enumerate}[(i)]
			\item \label{stmt:red:diff}
			If $\ell \geq \mygamma$, then
			$\confs(\ell, u) = \confs(\mydelta, \mydelta)
			+ \confs(\ell-\mydelta, u-\mydelta)$.
			\item \label{stmt:red:upper}
			If $u - \ell \geq \mygamma + \mydelta$, then
			$\confs(\ell, u) = 
			\confs(0, \mydelta) 
			+ \confs(\ell, u - \mydelta)$.
			\item {\label{stmt:red:lower}}
			If $\ell \geq \mygamma$, then
			$\confs(\ell, \infty) = \confs(\mydelta, \infty)
			+ \confs(\ell-\mydelta, \infty)$.
			\stopp
		\end{enumerate}
	\end{lemma}
	\input{proof___red___long.texb}
	By inductively applying this statement in an efficient manner, we get our desired compression results.
	In order to simplify the proof, we normalize $T$ to $1$ by scaling and rounding the speeds; afterwards for each machine $i$, the speed $s_{\resch(i)}$ and the capacity $T s_{\resch(i)}$ coincide.
	\begin{proposition}[Normalization]
	\label{stmt:normalize}
	\normalfont
		Consider a scheduling instance
		$I = (d,\allowbreak\tau,\allowbreak\vecn,\allowbreak\vecm,\allowbreak p,\allowbreak s)$ and a fractional number $T$.
		Then, in $\es$ time one can find 
		\begin{enumerate}[(i)]
			\item  an instance
				$I' \defeq (d, \tau, \vecn, \vecm, p, s')$ with $s' \in \N_0^d$ and
				$s'_{\max} \leq 1 + p^\intercal \vecn$ such that 
				$I$ is admits a schedule with makespan $\leq T$ iff $I'$ 
				admits a schedule with makespan $\leq 1$; and
			\item an instance
				$I' \defeq (d, \tau, \vecn, \vecm, p, s')$ with $s' \in \N_0^d$ and
				$s'_{\max} \geq 1 + p^\intercal \vecn$ such that 
				$I$ is admits a schedule with makespan $\leq T$ iff $I'$ 
				admits a schedule with makespan $\geq 1$.
			\end{enumerate}
			 \stopp
	\end{proposition}
	\begin{proof}
	We only prove the first case, the second one is analogously.

	First, note that it suffices to show that 
	in $\es$ we can find either such an $I'$ or conclude that
	$I$ does not admit a schedule with makespan $\leq T$.
	To see this, assume that $I$ does not admit such a schedule.
	We could then construct an instance $I'$ by simply setting the
	speed of all machines to $0$.
	Now assume that $I$ admits a schedule with makespan $\leq T$.
	Let $\sigma$ be such a schedule.
	For every machine $i \in \iv m$, we have
	$p^\intercal \sigma_i \leq s_{\resch(i)} \cdot T$.
	As the load $p^\intercal \sigma_i$ of machine $i$ is integral,
	we also have
	$p^\intercal \sigma_i \leq \floor{s_{\resch(i)} \cdot T}$.
	Note that we also have $p^\intercal \sigma_i
	\leq p^\intercal \vecn$.
	Thus, setting
	$s'_t \defeq \min\sing{
		\floor{s_{\resch(i)} \cdot T}
		,\,p^\intercal \vecn}$ for all $t \in \iv \tau$ 
	constructs an instance $I'$ with the desired property.
	\qed
	\end{proof}
	Eventually, we can prove our compression statement:
	\begin{lemma}[Compression]
		\normalfont
		\label{stmt:general_preprocessing}
		\label{stmt:compression}
		Consider a scheduling instance $I = (d, \tau, \vecm, \vecn, p, s)$.
		If considering \Cmin, let $(\rel) \coloneqq (\geq)$; otherwise, for \Cmax, let $(\rel) \coloneqq (\leq)$.
		and a guess $T$ of the optimal objective value. 
		Then, in time $\es$, one can find an instance
		$I' \defeq (d, \tau', \vecm', \vecn, p, s')$ such that
		\begin{enumerate}[(i)] 
			\item \label{stmt:general_preprocessing:feasible}\label{stmt:compression:feasible}
				$I$ admits a schedule where each machine $i \in \iv m$ has load $\rel s_{\resch(i)} T$ 
				iff $I'$ admits a schedule where each machine $i \in \iv m$ has load $\rel s_{\resch(i)}$\semi
			\item \label{stmt:general_preprocessing:smax}\label{stmt:compression:smax}
				The maximum machine speed is upper bounded, that is, $\smax' \leq \mygamma + \mydelta$\semi
			\item \label{stmt:general_preprocessing:tau}\label{stmt:compression:tau}
				The number of machine types increases by at most one and is upper bounded, that is, $\tau' \leq 1 + \min\sing{\enpar{\mygamma + \mydelta} ,\, \tau}$\semi&
			\item \label{stmt:general_preprocessing:my}\label{stmt:compression:my}
			The increase of the number of machines is bounded by the
			factor $\enpar{2 + p^\intercal \vecn}$, that is,
			$\norm {\vecm'}_1 \leq 
				\enpar{2 + p^\intercal \vecn}
				\cdot \norm {\mu}_1$\semi.
				\stopp
		\end{enumerate}
	\end{lemma}
	\input{proof___compression___long.texb}

	\section{Conversion}
	\label{sec:conversion}
	In this section, we formally show how to convert any \QCmin[\HM] instance with objective value $T$ to a \QCidle[\HM] instance with makespan $T + \pmax$ and idle load $\leq \pmax$, as well as any \QCmax[\HM] instance to a \QCidle[\HM] instance with makespan $T$ and idle load $0$.
	\begin{lemma} \label{stmt:conversion_cmin}
	\normalfont
	Given a number $T$, there is a schedule such that each machine $i \in \iv m$
	has load greater than $T \cdot s_{\resch(i)}$ (that is: no machine is finished before time $T$) iff there is a partial schedule such
	that the load of each machine at most than $T \cdot s_{\resch(i)} + \pmax$ and each machine
	has idle load less than $\pmax$.
	\stopp
	\end{lemma}
	\begin{proof}
		\enquote{$\Rightarrow$}
		Assume we are given a schedule $\sigma$ such that for all $i \in \iv m$ we have
		$p^\intercal \sigma_i \geq T \cdot s_{\resch(i)}$. Construct a partial schedule
		$\bar\sigma$ as follows: From each machine, remove jobs until any further removal of jobs
		would imply that the load of the machine drops below $T \cdot s_{\resch(i)}$.
		This implies that the load of each machine $i$ is now still $\geq T \cdot s_{\resch(i)}$
		but also $\leq T \cdot s_{\resch(i)} + \pmax$, as otherwise we could remove any remaining
		job without the load dropping below $T \cdot s_{\resch(i)}$.
		
		\enquote{$\Leftarrow$}
		Now assume that we are given a partial schedule $\bar\sigma$ such that the load
		of each machine is at most $T \cdot s_{\resch(i)} + \pmax$ and each machine has idle
		load less than $\pmax$. We may simply extend $\bar\sigma$ to a (non-partial) schedule
		by schedule the missing jobs arbitrarily. Doing so, we always end up with
		a feasible solution for the \Cmin\ decision problem, as every machine
		already has sufficiently load $\geq T \cdot s_{\resch(i)}$ under $\bar\sigma$ by definition of idle load, and we only
		further increased it by adding jobs.
		\qed
	\end{proof}
	Note that this implies that the makespan of the partial schedule is no larger than $T + \pmax$.
	\begin{lemma} \label{stmt:conversion_cmax}
	\normalfont
	Let $I$ be a \QCmax[\HM] instance. There is an instance $I'$ of \QCidle[\HM], s.t. there exists a schedule with makespan at most $T$ for \QCmax[\HM] iff there exists a schedule with makespan at most $T$ and idle load $\kappa = 0$.
	\stopp
	\end{lemma}
	\begin{proof}
		\enquote{$\Rightarrow$}
		We construct $I'$ from $I$ by introducing an appropriate amount of dummy jobs of processing time $1$ (precisely: $\sum_{i \in \iv m}\floor{s_{\resch(i)} T}$ many) to ensure that $I'$ has idle load exactly $0$ on each machine under makespan $T$, if a schedule with makespan $T$ exists.
		This construction is valid due to the fact that (i) this does not change feasibility, (ii) adds at most one new job type (iii) does not create too large numbers.
		The latter is important for applying this result algorithmically, because the running time of the algorithm of \textcite{DBLP:conf/soda/CslovjecsekEHRW21} does depend poly-logarithmically on the right hand side.
		However, this is not a problem: If $\sum_{i \in \iv m}\floor{s_{\resch(i)} T} \geq n^2 \pmax$, the instance can be solved in polynomial time.
		To see this, observe that by pigeonhole in any feasible solution, there is a machine that gets $m^{-1} (n+m)^2 \pmax \geq n\pmax$ dummy jobs, which means that placing all jobs (after removing dummy jobs from the instance) onto the fastest machine is a feasible solution.
		
		\enquote{$\Leftarrow$}
		As \QCidle[\HM] is a generalization of \QCmax[\HM], this follows immediately.
		\qed
	\end{proof}
\end{document}